\documentclass[useAMS,usenatbib]{mn2e}
\usepackage{soul}
\usepackage{epsfig}
\usepackage{multirow}
\usepackage{enumerate}
\usepackage{amsmath}
\usepackage{float}
\usepackage[section]{placeins}
\usepackage{color}

\title[FRB Classification Statistics...]
{Fast radio bursts: do repeaters and non-repeaters originate in statistically similar ensembles? }

\author[X.Cui et al.]
{Xiang-Han Cui$^{1,2}$, Cheng-Min
Zhang$^{1,2}\thanks{zhangcm@bao.ac.cn(CMZ)}$,
Shuang-Qiang Wang$^{3}$, Jian-Wei Zhang$^{4}$,
\newauthor  Di Li$^{1,2,5}$, Bo Peng$^{1,2,6}$, Wei-Wei Zhu$^{1,2}$, Na Wang$^{3}$, Richard  Strom$^{7,8}$,
\newauthor Chang-Qing Ye$^{9}$,  De-Hua Wang$^{10}$, Yi-Yan Yang$^{10}$\\
$^1$CAS key Laboratory of FAST, National Astronomical Observatories, Chinese Academy of Sciences, Beijing 100101, China\\
$^2$University of Chinese Academy of Sciences, Beijing 100049, China\\
$^3$Xinjiang Astronomical Observatory, Chinese Academy of Sciences, Urumqi, Xinjiang 830011, China\\
$^4$Department of Astronomy, Beijing Normal University, Beijing, 100875, China\\
$^5$NAOC-UKZN Computational Astrophysics Centre, University of KwaZulu-Natal, Durban 4000, South Africa\\
$^6$Key Laboratory of Radio Astronomy, Chinese Academy of Sciences, Beijing 100101, China\\
$^7$Netherlands Institute for Radio Astronomy (ASTRON), Postbus 2, 7990 AA Dwingeloo, the Netherlands \\
$^8$Astronomical Institute ‘Anton Pannekoek’, Faculty of Science, University of Amsterdam, the Netherlands\\
$^9$TianQin Research Center for Gravitational Physics, Sun Yat-sen University, Zhuhai 519082, China\\
$^{10}$School of Physics and Electronic Sciences, Guizhou Education University, Guiyang 550018, China\\
}
\date{Released 2020 July XX}

\pagerange{\pageref{firstpage}--\pageref{lastpage}} \pubyear{2002}

\begin{document}

\maketitle

\label{firstpage}

\begin{abstract}
Fast Radio Bursts (FRBs) are the short, strong radio pulses lasting several milliseconds.
They are subsequently identified, for the most part, as emanating from unknown objects at cosmological distances.
At present, over one hundred FRBs have been verified, classified into two groups:
repeating bursts (20 samples) and apparently non-repeating bursts (91 samples).
Their origins, however, are still hotly debated.
Here, we investigate  the statistical classifications for the two groups of
samples to see if the non-repeating and repeating FRBs have different origins by employing Anderson-Darling (A-D) test and Mann-Whitney-Wilcoxon (M-W-W) test. Firstly, by taking the pulse width as a statistical variant, we found that the repeating samples do not follow the  Gaussian statistics
(may belong to a $\chi$-square distribution), although the overall data and non-repeating
group do follow the Gaussian. Meanwhile, to investigate the
statistical differences between the two groups, we turn to M-W-W test
and notice that the two distributions have different origins.
Secondly, we consider the FRB radio luminosity as a statistical
variant, and find that both groups of samples can be regarded as
the Gaussian distributions under the A-D test, although they have
different origins according to M-W-W tests. Therefore,
statistically, we can conclude that our classifications of both repeaters and non-repeaters are plausible, that the two FRB classes have different origins, or each has experienced distinctive phases
or been subject to its own physical processes.
\end{abstract}


\section{Introduction}\label{1}

Fast radio bursts (FRBs) are bright, millisecond-duration radio
pulses that generated from the extragalactic sources (in most
cases) according to the dispersion measures (DM) and the redshift
of host galaxies of localized FRBs \citep{Lorimer07, Thornton13, Chatterjee17}.
Great progresses  have  been made in observations and theories,
since the first reported FRB in 2007
\citep{Cordes19,Petroff19}.
More than one hundred of FRBs have been verified, 20 (91) of which
are reported as repeating (apparently non-repeating) FRBs
\citep{Petroff16, CHIME19a, Kumar19, CHIME19b, Fonseca20}. The corresponding
FRB detection rate is $\sim 10^3-10^4 \space \rm{day^{-1}\space
sky^{-1}}$
\citep{Thornton13,Spitler14,Keane15,Rane16,Oppermann16,Champion16,Scholz16,Lawrence17,Patel18,Connor19}.
The physical origin of FRBs is still unclear, though
many theoretical models are proposed to solve this challenge
\citep{Cordes19,Petroff19}, including the  mergers of compact
objects \citep{Yamasaki18}, collapse of supermassive neutron stars
\citep{Falcke14}, energetic flares coming form the  magnetars
\citep{Kulkarni14,Connor16,Cordes16a,Popov16,Margalit18} and
interactions between superconducting cosmic strings
\citep{Yu14,Thompson17}.

Recently, bright millisecond-timescale radio bursts from the
magnetar SGR 1935 + 2154 have been detected by CHIME/FRB
\citep{Scholz20} and STARE2 \citep{Bochenek20a}. This phenomenon
suggests that some FRBs may be involved in the strong magnetic
activity generated by magnetars
\citep{Katz16,Margalit20,Andersen20,Lin20,Bochenek20b,Lyutikov20},
especially for the repeating FRBs. The repeaters
and apparently non-repeaters (hereafter referred to as non-repeaters) may have different physical origins. Thus, it is important to classify the FRBs based on the observed properties.

Clearly, it is natural to divide FRBs into two groups as
repeating and non-repeating samples \citep{Petroff19}. Both
pulse width and radio luminosity are important characteristics
for FRBs and their radiative properties, and these are frequently
used as FRB sorting criteria
\citep{Oppermann16,Ravi19,Qiu20,Fonseca20}. \citet{Petroff19}
presented the histogram of FRB pulse width, however
further statistical tests were not pursued. Recent results from
the CHIME/FRB collaboration show that the distributions of two
categories are not same with \rm{$\sim 5\sigma$ and $\sim
4\sigma$} significance using analysis of their own data
on 18 repeating FRBs and 12 non-repeating ones \citep{Fonseca20}. The ASKAP
group found that the distribution of pulse width may not be
bimodal \citep{Qiu20}. But considering that the ASKAP sample is small and Bayesian  methods are more suitable for large samples, the conclusion is still inconclusive.
Meanwhile, much research on the radio luminosity of FRBs have been carried out, which mainly focuses on
the luminosity function \citep{Kumar17, Luo18, Luo20, Hashimoto20}
, and radio spectrum \citep{Spitler16, Macquart19, Katz20}. Although the
above analyses are impressive, the statistics combined with more
data from other telescopes and quantitative statistical tests are
still needed.

Here, we collect all detected FRB data, including the
information on pulse width and radio luminosity, to examine their
distributions for the repeating and non-repeating samples and
check whether two groups of samples have the same or different origins.
In section 2, we organize and check the rationality of
the data. In section 3, we exploit the  Anderson-Darling (A-D) test
to see whether the above data conform to the Gaussian distribution
and use a Mann-Whitney-Wilcoxon (M-W-W) test to check whether the
two distributions share  the same origins. Finally, in section 4
we  summarize our results, and discuss  possible mechanisms
for the  repeating and non-repeating FRBs.

\section{FRB Observation Data}\label{2}

Our data of FRBs are taken  from the database of FRB
Catalogue\footnote{http://www.frbcat.org/}(FRBCAT)
\citep{Petroff16}, and those for repeating FRBs come from
FRBCAT and published papers \citep{Petroff16, CHIME19a, Kumar19, CHIME19b,Fonseca20}.
According to FRBCAT, there is a significant gap between $\sim35\,\rm{ms}$ and $\sim300\,\rm{ms}$
in pulse width. However,  we notice that the measurements with
pulse width larger than $\sim300\,\rm{ms}$ all come from one radio
facility, i.e., the radio telescope BSA LPI of the Pushchino Radio
Astronomy Observatory, in  which the time interval between
samples is $100\,\rm{ms}$ \citep{Fedorova19}. Considering that
there may exist  some observational bias  effects in these data,
we only include the pulses with a width shorter than
$\sim35\,\rm{ms}$ (80 non-repeaters).
For non-repeating FRBs, the possibility of repetition cannot be rejected, in
particular repeating signals from FRB171019 have been observed \citep{Kumar19} in 2019.
However, our analysis is based on the current observational data, and the phenomenon of repeating signals in non-repeaters is difficult to predict at present.
Therefore, we treat the apparently non-repeating sources as real non-repeaters under current circumstances.
For repeating FRBs, two or more observations are included. Thus, we calculate the average pulse
width and radio luminosity of each source as representative of its pulse
width and radio luminosity, respectively.
Since the pulse width in the FRBCAT is the observed width, which is easily affected by dispersion\citep{Ravi19}, to study the pulse width more accurately, we need to introduce the intrinsic width that is estimated by Eq.(1) \citep{Connor20, Qiu20}.

\begin{equation}
\begin{split}
t_i = \sqrt {t_{obs}^2 - t_{DM}^2 - t_s^2}
\end{split}
\end{equation}
In the above formula, $t_i$ ($t_{obs}$) is the intrinsic width (observed width), with  $t_s$  being  the
 sampling time that depends on the instrument,  and $t_{DM}$ is the dispersion smearing timescale as  calculated in the following,

\begin{equation}
\begin{split}
t_{DM} = 8.3 \times 10^{-3} \rm{DM {\Delta {\nu_{MHz}}\over\nu_{GHz}^3} \quad ms}
\end{split}
\end{equation}
where $\rm DM$ is the dispersion measure, $\rm\Delta {\nu_{ MHz}}$ is the channel bandwidth in the unit of
MHz and $\nu_{\rm GHz}$ is the central frequency in the unit of GHz.
Therefore, the pulse width in the following text represents the intrinsic width.
In Table 1, we summarize
the properties of 20 repeating FRBs, by listing the observed width, intrinsic width,
flux density, fluence and luminosity distance. In Figure 1, we plot
a histogram of the pulse width for the repeating and
non-repeating FRBs.

\begin{figure}
\centering
\includegraphics[width=7.8cm]{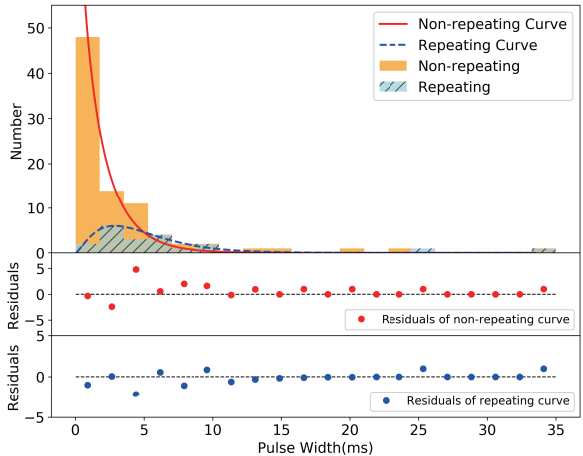}
\caption{Upper panel: histogram of repeating and non-repeating FRBs with pulse
width $<35 \, \rm {ms}$. The solid (dashed) line is the fitted curve
for non-repeating (repeating) FRBs.  The filled histogram is for
non-repeating FRBs and the  cross-hatched histogram is for
repeating FRBs.
Middle and bottom panels: the residuals for the  non-repeaters and repeaters between the source data and  fitted curve,
shown as solid points, where the dashed lines refer to the cases of residual=0.}
\label{fig1}
\end{figure}

\section{FRB  Classifications and  Statistical Tests}\label{3}

\subsection{Samples of all FRBs}

A Gaussian distribution occurs naturally in many astronomical
data sets \citep{Press86, Mackay03},
sometimes occurring in data with logarithmic
sampling. For completeness,
we first investigate the Gaussian property.
Regarding the total statistical properties of all FRBs, if we
assume that the pulse widths of these data follow the Gaussian
distribution, we can apply the A-D test \citep{Ivezi19} to
examine whether the assumption is correct. The results of A-D test
are shown in Table 2. For a 0.05 significance level, the statistic
(0.547) is smaller than the critical value (0.760), which indicates
that the Gaussian  property of the data on pulse width is probable. We further use the A-D test to examine whether the radio
luminosity of FRBs conforms to a Gaussian distribution, and the result
shows that its statistic (0.475) is smaller than the critical
value (0.759, as seen in Table 2), which means that the Gaussian
property of the radio luminosity is likely. Therefore, using a
Gaussian distribution to fit for the pulse width and radio luminosity is acceptable for all data but unknown for the separate data.
Next we will discuss the distributions for the  two samples of repeaters and non-repeaters.

\begin{table*}
\centering
\caption{Parameters of 20 repeating FRBs. $^a$}
\resizebox{\textwidth}{!}{
\begin{tabular}{@{}lccccccc@{}}
\hline
\noalign{\smallskip}
\bf No.&\bf Sources &\bf Observed Width$^b$ &\bf Intrinsic Width$^c$&\bf Flux Density$^d$&\bf Fluence$^e$&\bf Distance$^f$& \bf Refs. \\
\noalign{\smallskip}
\bf & &\bf (ms) &\bf (ms)&\bf (Jy)&\bf (Jy ms)&\bf (Gpc) &  \\
\hline
\noalign{\smallskip}
1&FRB121102 & 4.82 &4.78& 0.25 &0.372 & 1.61 & (1)\\
2&FRB180814.J0422+73 & 23.45(22.57)$^g$  &23.43&...$^h$ & 22.57 & 0.39 & (2)(3)\\
3&FRB171019 & 4.62 &4.08&  ...$^{hi}$ & 101.54 & 1.89 & (4)\\
4&FRB180916.J0158+65 & 5.27 &5.16& 2.08 & 1.62 & 0.58 & (5)\\
5&FRB181030.J1054+73 & 1.01 &0.10& 3.15& 4.75 & 0.24 & (5)\\
6&FRB181128.J0456+63 & 5.90 &5.80& 0.40& 3.45 & 1.14 & (5)\\
7&FRB181119.J12+65 & 3.49 &3.33& 0.43& 1.77 & 1.42 & (5)\\
8&FRB190116.J1249+27 & 2.75 &2.53&0.35 & 1.80 & 1.90 & (5)\\
9&FRB181017.J1705+68 & 16.8 &16.73&0.40 & 8.50 & 6.97 & (5)\\
10&FRB190209.J0937+77 & 6.55 &6.46&0.50 & 1.25 & 1.66 & (5)\\
11&FRB190222.J2052+69 & 2.71 &2.48&1.65 & 5.45 & 1.64 & (5)\\
12&FRB190208.J1855+46 & 1.11 &0.14& 0.50 & 1.70& 2.35 & (6)\\
13&FRB180908.J1232+74 & 3.83 &3.70& 2.90 & 0.50 & 0.62 & (6)\\
14&FRB190604.J1435+53 & 2.10 &1.78& 0.75 & 8.30& 2.42 & (6)\\
15&FRB190212.J18+81 & 3.10 &2.93& 0.75 & 2.75& 1.05 & (6)\\
16&FRB190303.J1353+48 & 3.20 &3.04& 0.47 & 2.67& 0.77 & (6)\\
17&FRB190417.J1939+59 & 4.50 &4.20& 0.53 & 3.10& 7.40 & (6)\\
18&FRB190117.J2207+17 & 2.74 &2.53& 1.00 & 6.36& 1.49 & (6)\\
19&FRB190213.J02+20 & 7.00 &6.90& 0.50 & 1.80& 2.91 & (6)\\
20&FRB190907.J08+46 & 2.18 &1.92& 0.30 & 2.03& 1.07 & (6)\\
\hline
\end{tabular}
}
\label{tab1}
\begin{flushleft}
$^a$ The highest radio luminosity is $1.2\times 10^{44}\,\rm{erg/s}$ from FRB190523, and the faintest for extragalactic conditions is $6.2\times 10^{38}\,\rm{erg/s}$ from FRB141113.
$^b$ The average observed width.
$^c$ The average intrinsic width calculated by eq.(1).
$^d$ The average flux density.
$^e$ The average fluence.
$^f$ The distance is from FRBCAT \citep{Wright06} based on $\rm {\Lambda CDM}$ with cosmological parameters: $\rm{H_0=69.6\,km/s/Mpc}$, $\rm{\Omega_M=0.286}$ and $\rm{\Omega_{vac}=0.714}$,
where $\rm{H_0}$ is Hubble constant, $\rm{\Omega_M}$ is the mass fraction in the universe and $\rm{\Omega_{vac}}$ is the dark energy fraction in the universe.
$^g$ 23.45 is from CHIME \citep{CHIME19a}, while 22.57 is from FRBCAT. In our statistics we use the data from CHIME.
$^h$ The parameters are not given in FRBCAT or Refs..
$^i$ The flux density is not given, but the fluence is given. So we use the fluence to estimate the luminosity.
\\
Refs. : (1) Spitler et al. (2016);
(2) CHIME/FRB Collaboration et al. (2019a);
(3) FRBCAT;
(4) Kumar et al. (2019);
(5) CHIME/FRB Collaboration et al. (2019b);
(6) Fonseca et al. (2020).
\end{flushleft}
\end{table*}


\subsection{Samples of repeating and non-repeating FRBs}
To obtain the statistical test results for FRB classifications,
we can apply the A-D test and M-W-W test.
Here, due to the small size of repeater sample (20 repeating FRBs),
a Kolmogorov-Smirnov (K-S) test is not an effective tool \citep{Ivezi19, Yang19}.
For example, for a sample size of 10, the error of a K-S test can reach 7\%\footnote{https://en.wikipedia.org/wiki/Kolmogorov-Smirnov\_test}.
Thus, we adopt the A-D test to check whether the sample distribution is consistent with a single Gaussian,
then use the M-W-W test to examine whether the two samples share
the same origin even if the data volumes are different.
In our case, the two groups of samples for repeating and
non-repeating FRBs are tested, based on  the two characteristic
quantities, the pulse width and radio luminosity. On the one hand,
according to the data in Table 1, there are 20 repeaters,
the pulse width of which range from $\sim0.1\,\rm{ms}$ to $\sim23\,\rm{ms}$.
On the other hand, for the non-repeating FRB data, the pulse widths
range from $\sim0.05\,\rm{ms}$ to $\sim34\,\rm{ms}$.
The mean pulse width of repeating (non-repeating) sources is
$\sim5.10\,\rm{ms}$ ($\sim3.35\,\rm{ms}$). We use A-D test
to check the Gaussian property of both groups of samples, the results
of which are shown in Table 2.

The results show that the
distribution of pulse width for repeating samples is not Gaussian in either logarithmic or linear coordinates scales, while the non-repeating sample belongs
to a Gaussian distribution if expressed logarithmically.
Furthermore, to check if the distribution of repeating FRB pulse widths
follows a $\chi$-square type in linear coordinates, the function of which can be described
below,

\begin{equation}
\begin{split}
f_{k}(x)&={1\over2^{{k/2}}\Gamma({k/2})}x^{k/2-1}e^{-x/2}\\
&=Ax^{k/2-1}e^{-x/2}\
\end{split}
\end{equation}
where A is referred as the fitting coefficient.
Here we employ  Eq. (3) to fit the pulse width data of repeaters with the best fitting parameters of $\rm A=72.978$ and $\rm k=1.626$.
The goodness of the $\chi$-square fit is calculated by Eq. (4) and the result is 0.80.
In Eq. (4), $R^2$ is the goodness of fit, RSS is residual sum of the square, TSS is the total sum of squares, $y_i$ values are real data, $\hat y$ are test data on the fitted curve and $\overline y$ is mean value of the real data.
Through the Figure 1, the histograms are all on the fitted curve or on both sides of the curve.
Thus, we conclude that  the distribution of pulse widths of repeating FRBs could follow a $\chi$-square function.

\begin{equation}
\begin{split}
R^2&=1-RSS/TSS\\
&=1-\sum\limits_{i=1}^{n}(y_i-\hat y)^2/\sum\limits_{i=1}^{n}(y_i-\overline y)^2
\end{split}
\end{equation}
In Figure2, we draw the cumulative distribution function (CDF) for the two
groups of samples. Although the two CDFs are close to each other,
they belong to different distributions. To evaluate the
reliability of above inference, we apply a M-W-W test. The resulting
p-value of this M-W-W test is 0.0065, which is less  than 0.05,
indicating that the distribution of two groups are different, as
shown in Table 3.

As a next step,  we try to use the radio luminosity as a
statistical variant to realize classification for two group
samples. The radio luminosity is estimated by Eq. (5), where S is
the flux density and D is the luminosity distance.

\begin{figure}
\centering
\includegraphics[width=7.8cm]{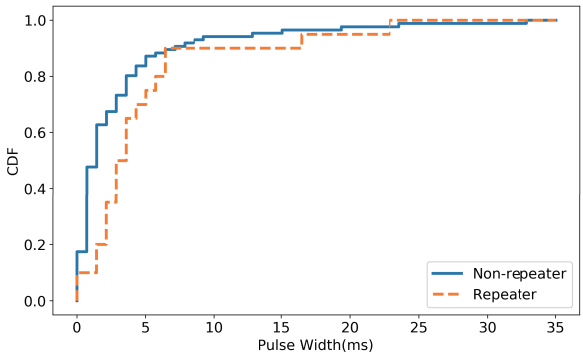}
\caption{Cumulative distribution function (CDF) of pulse width.
The dashed line is for repeating FRBs, the solid line is for
non-repeating FRBs.}
\label{fig2}
\end{figure}

\begin{figure}
\centering
\includegraphics[width=7.8cm]{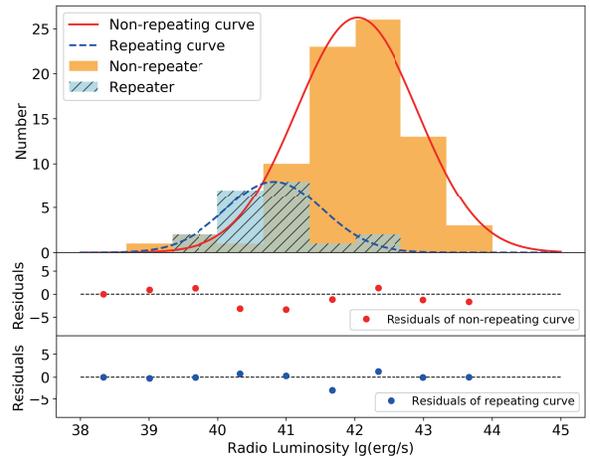}
\caption{Histogram of repeating and non-repeating FRBs for radio luminosity  expressed logarithmically.
The solid line is the fitted curve of non-repeating FRBs.
The dashed line is the fitted curve of repeating FRBs.
The cross-hatched histogram is repeating FRBs and the empty one means non-repeating FRBs.
Middle and bottom panels: the residuals for the non-repeaters and repeaters between the  source data and  fitted curve,
shown as solid points, where the dashed lines refer to the cases of residual=0.}
\label{fig3}
\end{figure}

\begin{figure}
\centering
\includegraphics[width=7.8cm]{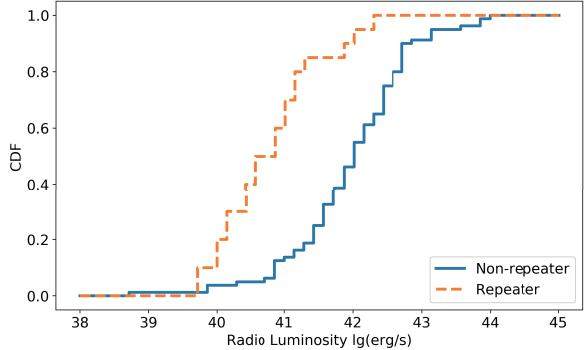}
\caption{Cumulative distribution function (CDF) of radio luminosity.
The dashed line is for repeating FRBs and the soild one shows the non-repeating FRBs.}
\label{fig4}
\end{figure}

\begin{equation}
\begin{split}
L_{\rm{radio}}\sim SD^2\
\end{split}
\end{equation}
First of all, the radio luminosity of repeaters and non-repeaters
ranges from $\sim10^{39}\,\rm{erg/s}$ to $\sim10^{42}\,\rm{erg/s}$
and $\sim10^{38}\,\rm{erg/s}$ to $\sim10^{44}\,\rm{erg/s}$,
respectively. We plot their  histograms  in Figure 3. The mean
radio luminosity of repeating (non-repeating) sources is
$\sim2.6\times 10^{41}\,\rm{erg/s}$ ($\sim6.2\times
10^{42}\,\rm{erg/s}$). Then, we need to test for the Gaussian property
of  the repeating and non-repeating groups by using A-D test, the
results of which are shown in Table 2:  for the repeating FRBs the
statistic (0.333) is smaller than the critical value (0.692), and
for the non-repeating FRBs the statistic (0.592) is also smaller
than the critical value (0.752), so they are both under the
corresponding significance level of 0.05. Hence, these results
indicate that the distributions of two groups conform to Gaussian distributions.
In Figure 4, we
draw the CDF for two groups, finding that the two curves are well-separated.
We find both distributions are different using the M-W-W
test, for which p-value is $7.905\times 10^{-7}$, much smaller
than 0.05, as shown in Table 3. In short, from the viewpoint of
statistical tests, two groups of samples very likely have different
distributions.

\begin{table}
\centering
\caption{A-D test to check Gaussian property $^a$.}
\begin{tabular}{@{}lccc@{}}
\hline
\noalign{\smallskip}
\bf Sample &\bf Statistic $^b$ &\bf Critical values ($\alpha=0.05$)$^c$ \\
\hline
\noalign{\smallskip}
\bf Pulse Width & &\\
All Data & 0.547 & 0.760 \\
Non-repeating & 0.692 & 0.754 \\
Repeating & 1.430 & 0.692 \\
\bf Luminosity & &\\
All Data& 0.475 & 0.759 \\
Non-repeating & 0.592 & 0.752 \\
Repeating & 0.333 & 0.692 \\
\hline
\end{tabular}
\label{tab3}
\begin{flushleft}
$^a$ Tested when in logarithmic scale.\\
$^b$ The statistic in the A-D test. Only when it is smaller than a critical value does the result make sense;\\
$^c$ The significance level in A-D test we take to be 0.05.\\
\end{flushleft}
\end{table}

\begin{table}
\centering
\caption{M-W-W test to check different distributions.}
\begin{tabular}{@{}lcccccc@{}}
\hline
\noalign{\smallskip}
\bf Characteristic  &&&&&\bf P-values \\
\hline
\noalign{\smallskip}
Pulse Width &&&&&  0.0065 \\
Radio Luminosity &&&&&   $7.905\times10^{-7}$\\
Radio Luminosity in CHIME Data  &&&&&   $8.79 \times 10^{-6}$\\
Adjusted Radio Luminosity &&&&&   $2.47 \times 10^{-9}$\\
\hline
\end{tabular}
\label{tab3}
\end{table}

However, we notice that the mean values of radio luminosity for non-repeaters are different for the CHIME data with central frequency of 600MHz ($\rm{1.77 \times 10^{43} \, erg/s}$) and other data centered  around 1.4GHz ($\rm {2.59 \times 10^{42} \, erg/s}$) \citep{Luo20}.
If we assume that the detected FRBs by both the CHIME and other radio telescopes are originated from the same phenomenon,
  then the two mean values of  luminosity should be similar in the case of readjusting the observational frequency.
  In other words, the two different  mean values of luminosity of FRBs by the different instruments  may be caused by the observation frequency bands or facilities calibration.
First, we simply use the CHIME data \citep{Fonseca20} with 18 repeaters and 12 non-repeaters to test the former conclusion, whether the two samples have the same distributions. Through M-W-W test, the p-value is $8.79 \times 10^{-6}$ which is consistent with the former conclusion that they follow
the different origins statistically.
Second,  we shift  the luminosity values of CHIME  by reducing an order of magnitude, which results in the same mean value as that by
the other telescopes,  as shown in Figure 5.
Obviously, the two samples of the adjusted non-repeaters and repeaters  are tested  by applying M-W-W test, as shown in Table 3, and the result
shows  the same conclusion as the former  that the repeaters and non-repeaters belong to the different distributions.

\begin{figure}
\centering
\includegraphics[width=7.8cm]{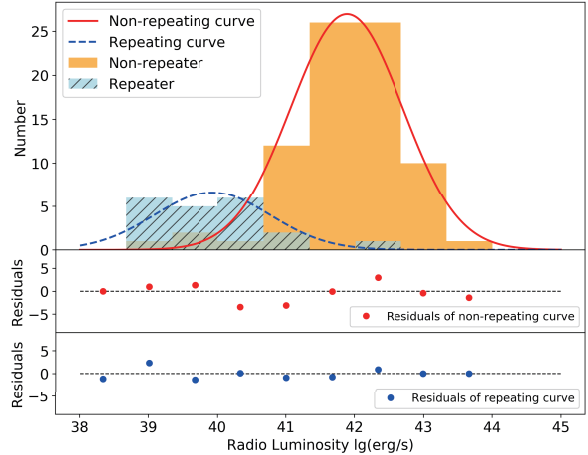}
\caption{ Histogram of repeating and non-repeating FRBs for radio luminosity on a logarithmic scale, after the
adjustment of data.
The solid (dashed) line stands for  the fitted curve of non-repeating (repeating) FRBs.
The cross-hatched histogram is for repeating FRBs and the empty histogram means the non-repeating ones.
Middle and bottom panels: residuals of non-repeaters and repeaters between source data and the fitted curve,
shown as solid points. Dashed lines refer to residual=0.}
\label{fig5}
\end{figure}

\section{Conclusions and Discussions}\label{4}
In this paper, we employ statistical methods to analyze the distributions of FRB properties based on the statistical variants of
FRB pulse width and radio luminosity, in an attempt to classify the two sample groups, the repeating (20
samples) and non-repeating (80 samples) bursts.
Because of the small FRB sample sets at present, we avoid the K-S test and employ the M-W-W test due to larger uncertainties of the former.
We find that the two groups of samples have different origins, with the significance level of 0.05.

We think that the statistical classification turns out to be an effective guide to understanding FRB origins.
Firstly, taking the pulse width as a statistical variant, the distribution of non-repeating group is conforms to a Gaussian, but a
Gaussian property for the repeating group is not clear ($\chi$-square distribution seems to be better) when an A-D test is applied.
Furthermore, using M-W-W test, we find that the distributions of two groups, repeating and non-repeating, are different.
Secondly, in terms of radio luminosity, by adopting the A-D test, we find that the distributions of repeating and non-repeating groups are both Gaussian. A further  M-W-W test shows that the distributions of two groups are significantly different.
In addition,  for a more complete conclusion, we notice  that  the  CHIME data present the FRB luminosity of non-repeaters
one order of magnitude higher  than that of the other telescopes, in average.  Again, we adjust the CHIME data by reducing one order of magnitude
to test if the modified data of CHIME repeaters has the different origin from the non-repeaters, and we obtained the same
conclusion as that for the unadjusted data by M-W-W test: non-repeaters and repeaters have the different origins.
Therefore the statistical difference between two samples of data indicates that they may have a different physical origin, or the repeating and non-repeating phenomena originates in very different physical processes.

There are some interpretations concerning our data and conclusion which need to be clarified.
1) The pulse width data from FRBCAT is the observed value and further obtained by search code \citep{Petroff16}, but it is not the intrinsic width.
Moreover, according to papers published by CHIME \citep{CHIME19b} and Kumar et al. (2019), the data they present have been fitted with Gaussian profiles, which means that these data are the observed pulse width.
Since the main purpose of this paper is not to discuss the DM-t relationship, we directly use the Eq.(1) to estimate the intrinsic width.
Thus, for these data, we do not apply further processing to remove the  additional intrinsic structure effects.
2) The pulse width and flux density data are not corrected precisely for frequency, because the spectral index of each FRB is difficult to determine, in particular for the lack of data for simultaneous observation of the same source at different frequencies.
3) The distance we use is from FRBCAT \citep{Wright06} based on $\rm {\Lambda CDM}$ with cosmological parameters: $\rm{H_0=69.6\,km/s/Mpc}$, $\rm{\Omega_M=0.286}$ and $\rm{\Omega_{vac}=0.714}$. Some papers on luminosity function, such as the work by  Luo et al.\citep{Luo20}, use $\rm {\Lambda CDM}$ with cosmological parameters: $\rm{H_0=67.8\,km/s/Mpc}$, $\rm{\Omega_M=0.308}$ and $\rm{\Omega_{vac}=0.692}$.
4) The goodness of repeaters' pulse width fitting curve is only 0.80, which is not high enough.
This indicates  that even though the distribution of repeaters' pulse width is not the Gaussian, the $\chi$-square type should be not the best.
The explanation for this  may be due to the lack of observation data.

To explain the physical origins of repeating and non-repeating FRB mechanisms,
many FRB models have been proposed.
Now that we find two "congenital" origins related to different physical processes,
the two camps of models will be briefly summarized.
In general, FRB models usually involve the physical processes of compact objects [white dwarf (WD), neutron star (NS), black hole (BH)], or the medium around them.
Repeating FRB models usually relate to the interaction of compact objects with companion stars or the surrounding medium, including the super giant pulses from young NS or pulsars \citep{Cordes16a, Connor16}, extreme activities in the magnetosphere magnetars \citep{Kulkarni14, Wang20, Lyutikov20},
accretion of compact objects \citep{Gu16}, NS interaction with comets or asteroids \citep{Dai16}, and maser phenomenon in the surrounding medium of magnetars \citep{Yu20}.
For the non-repeating FRBs, many models involve the one-off explosive events such as the mergers of compact object binaries or collisions  between one  compact object and another astronomical object, for example, mergers of BH-NS \citep{Zhang16}, NS-NS  \citep{Yamasaki18}, WD-BH \citep{Li18}, or comets and asteroid hitting the surface of a NS \citep{Geng15}.

Because our statistical results support the classification of FRB into repeating and non-repeating categories, our work puts certain constraints on the different models.
For example, a model needs to be discussed from the perspective of repeating and non-repeating,
especially the luminosity difference between repeaters and non-repeaters is about 1.5 orders of magnitude. This difference may indicate that the non-repeating sources come from a onetime catastrophic energy release or a violent outburst with a long energy storage period.

Furthermore, compared with the latest results from CHIME \citep{Fonseca20} and Luo et al.\citep{Luo20}, all data in FRBCAT have been considered and the new method of M-W-W test has been used in this work.
However, we have to note that different observed center frequencies may cause the changes in pulse width and radio luminosity, which may be the reason why the former research did not consider all the data.
Finally, long-term observations of the repeating sources will test the different models and give us a better understanding of their burst mechanisms.
We hope that in terms of further observations, the same FRB can be observed at the  different frequency bands simultaneously.
In this way, we can more accurately determine the spectral index of FRBs, thereby constraining the luminosity function of FRBs.
Many more FRBs are expected to be published soon from CHIME, ASKAP, and FAST \citep{Li18, Li19, Zhu20},
the higher sensitivity of which could provide valuable information regarding their still mysterious origin.

\section*{Acknowledgments}

This work is supported by the National Natural Science Foundation of China (Grant No.U1938117, No. 11988101, No. U1731238 and No. 11703003), the International Partnership Program of Chinese Academy of Sciences grant No. 114A11KYSB20160008, the National Key R\&D Program of China No. 2016YFA0400702, and the Guizhou Provincial Science and Technology Foundation (Grant No. [2020]1Y019).
 And, we thank the anonymous referee especially for the critical comments and suggestions, which have significantly improved the quality of the paper.

\section*{Data Availability}

The data underlying this article are available in the references below: (1) Spitler, et al. (2016); (2) CHIME/FRB Collaboration, et al. (2019a); {\bf (3) Kumar et al. (2019);}
(4) CHIME/FRB Collaboration, et al. (2019b); (5) Fonseca, et al. (2020). Some data of FRBs are taken from the database of FRB Catalogue (FRBCAT), available  at http://www.frbcat.org/.

\bsp

\label{lastpage}

\end{document}